\documentclass[12pt,english]{article}
\usepackage[T1]{fontenc}
\usepackage[latin9]{inputenc}
\usepackage{geometry}
\usepackage{mathrsfs}
\usepackage{amsmath}
\usepackage{amssymb}
\usepackage{graphicx}
\usepackage{babel}
\begin{document}
\title{Critical Phenomena of Single and Double Polymer Strands in a Solution}
\author{R. Dengler\thanks{ORCID: 0000-0001-6706-8550}}
\maketitle
\begin{abstract}
A universality class describing the statistics of the merging of two
single polymer strands to a double polymer strand and the reverse
process is examined. The polymers can have an intrinsic direction,
and the simpler case, where only single strands aligned parallel bind
to a double strand is considered in detail. The critical dimension
of the universality class is six, there is a stable fixed point and
critical exponents are calculated with renormalization group and loop
expansion. The corresponding field theory describes polymer configurations
in terms of fields and not in terms of coordinate paths and is second
quantized in this sense.
\end{abstract}

\subsection*{Introduction }

It is known since a long time that the configurations of a long polymer
chain in a solution exhibit scaling behavior, equivalent to the $n\rightarrow0$
limit of the $O\left(n\right)$-symmetric Heisenberg magnet universality
class. Such systems can be described with a (``first quantized'')
path integral over the monomer coordinates, or with the $O\left(n\right)$
symmetric $\left(\mathbf{\varphi}^{2}\right)^{2}$ field theory.\cite{DeGen72}
\cite{Edwards1965}\cite{Cloiz81}\cite{ZinnJustin} The difficult
part of the problem is ``self avoidance'', the fact that a polymer
chain cannot intersect itself.

It also is known since a long time that the formation of a DNA double
strand from DNA single strands has aspects of first order phase transitions
and second order phase transitions. For instance, it is found to sometimes
proceed in steps and sometimes continuously when the temperature is
lowered. The real problem of course is complicated by the quasi irregular
nucleotide sequences, but a simple argument shows that the problem
is nontrivial. An estimate for the configuration entropy of a single
polymer strand of length $L$ in $d$ dimensions (ignoring self avoidance)
is $k_{B}\ln\left(2d-1\right)^{L}=k_{B}L\ln\left(2d-1\right).$ A
single strand bound to a double strand has half the length and half
the configuration entropy. This can be compensated by the binding
energy which also is proportional to $L$. 

Theoretical models mostly have considered polymer strands with a uniform
binding energy, and many methods have been used, see for example ref.\cite{KMP00}
A recent review is\cite{VK2017}. In this work we derive a field theory
from a microscopic model, which then allows to examine the statistical
mechanics of the binding and unbinding of double strands with the
renormalization group. 

\subsection*{The model}

We consider directed polymer chains (i.e., chains with an intrinsic
direction) and different binding energies for double strands consisting
of single strands aligned parallel and for double strands consisting
of single strands aligned antiparallel. This could be realized with
DNA chains. Assume there are nucleotide pairs (Xx) and $(Yy)$. Then
a strand ...XYxyXYxy... binds to itself with offset 2, but binds less
strongly to the inverted strand ...yxYXyxYX.... The model of course
also contains the case of equal binding energies. Similarly the intrinsic
direction also contains the case of undirected polymers, the direction
then only is a technical device.

The model, as appropriate for critical phenomena, consists of a path
integral with action

\begin{align}
S & =\int d^{d}x\left\{ \mathfrak{\mathcal{L}}_{0}+\mathcal{L}_{1}+\hat{\mathcal{L}}_{0}+\hat{\mathcal{L}}_{1}\right\} ,\nonumber \\
\mathfrak{\mathcal{L}}_{0} & =-\int ds\tilde{\varphi}\left(s\right)\left(r_{0}-\nabla^{2}+\partial_{s}\right)\varphi\left(s\right)-\int dsds'\tilde{\psi}\left(s,s'\right)\left(\tau_{0}-\nabla^{2}+w\left(\partial_{s}+\partial_{s'}\right)\right)\psi\left(s,s'\right),\nonumber \\
\mathcal{L}_{1} & =\frac{g}{\sqrt{K_{d}}}\int dsds'\left\{ \tilde{\psi}\left(s,s'\right)\varphi\left(s\right)\varphi\left(s'\right)+\tilde{\varphi}\left(s\right)\tilde{\varphi}\left(s'\right)\psi\left(s,s'\right)\right\} ,\label{eq:FieldTheory}\\
\hat{\mathcal{L}}_{0} & =-\int dsds'\tilde{\chi}\left(s,s'\right)\left(\tau_{0}-\nabla^{2}+\hat{w}\left(\partial_{s}-\partial_{s'}\right)\right)\chi\left(s,s'\right),\nonumber \\
\mathcal{\hat{L}}_{1} & =\frac{\hat{g}}{\sqrt{K_{d}}}\int dsds'\left\{ \tilde{\varphi}\left(s\right)\varphi\left(s'\right)\left(\chi\left(s,s'\right)+\tilde{\chi}\left(s',s\right)\right)\right\} .\nonumber 
\end{align}
For notational simplicity here all space arguments of fields are suppressed,
$\tilde{\varphi}\left(s\right)$ should be read as $\tilde{\varphi}\left(\mathbf{x},s\right)$
etc. The constant $K_{d}$ is defined in appendix B. A derivation
of the field theory with the help of operators on a lattice can be
found in appendix A. However, all terms have a clear meaning.

The field $\tilde{\varphi}\left(\mathbf{x},s\right)$ creates a start
of a single strand at position $\mathbf{x}$ with initial length variable
$s$, the field $\varphi\left(\mathbf{x},s\right)$ terminates a single
strand with length variable $s$ at position $\mathbf{x}$. The $\varphi$
propagator ($\mathfrak{\mathcal{L}}_{0}$, with $r_{0}=0$) describes
diffusion of the $\varphi$-field in $s$-direction (this is not a
diffusion that takes place step by step in $s$, it simply is part
of the partition sum).

The nonlinear terms in $\text{\ensuremath{\mathcal{L}_{1}}}$ can
be identified as the transition of two aligned single strands $\varphi$
to an (aligned) double strand $\tilde{\psi}\left(\mathbf{x},s,s'\right)$
and the reverse process. The terms in $\hat{\mathcal{L}}_{1}$ can
be identified as the transition of two oppositely aligned single strands
$\varphi,\tilde{\varphi}$ to an (oppositely aligned) double strand
$\tilde{\chi}\left(\mathbf{x},s,s'\right)$ and the reverse process.
The double strands $\psi$ and $\chi$ take over the length variables
of the single strands. Along a double strand of course one has $s-s'=const$
or $s+s'=const$, but field theory and perturbation theory are simpler
when both length variables are kept. One might introduce fields like
$\psi_{a}\left(\mathbf{x},s\right)$, where $a$ is the length offset
of the single strands. But then the interaction terms would be more
complicated.

The interactions $\mathcal{L}_{1}$ and $\hat{\mathcal{L}}_{1}$ are
local in space but non-local and even translationally invariant in
the length variables. This means that the interaction of two strands
at the same point in space is the same, for any length index. Technically
this implies that in Fourier space (with ``frequencies'' $\omega$
and $\omega'$ dual to $s$ and $s'$) no frequency is transferred
between single strands, even when the strands combine to a double
strand and separate again - the perturbation theory does not contain
any frequency integrals.

The redundancy (or gauge invariance) of length variables in the double
strand fields $\psi,\tilde{\psi},\chi,\tilde{\chi}$ reflects itself
in their propagators. In wavevector space the harmonic equations of
motion (functional derivative of $\mathfrak{\mathcal{L}}_{0}$ and
$\mathfrak{\mathcal{\hat{L}}}_{0}$ with respect to $\tilde{\psi}$
or $\tilde{\chi}$) read (the quantities $w$ and $\hat{w}$ are constants)
\begin{align*}
\left(\tau_{0}+\mathbf{k}^{2}+w\left(\partial_{s}+\partial_{s'}\right)\right)\psi_{\mathbf{k}}\left(s,s'\right) & =0,\\
\left(\hat{\tau}_{0}+\mathbf{k}^{2}+\hat{w}\left(\partial_{s}-\partial_{s'}\right)\right)\chi_{\mathbf{k}}\left(s,s'\right) & =0.
\end{align*}
The generic solutions of these homogeneous equations are
\begin{align*}
\psi_{\mathbf{k}}\left(s,s'\right) & =\exp\left(-\tfrac{1}{2w}\left(\tau_{0}+\mathbf{k}^{2}\right)\left(s+s'\right)\right)\gamma_{\mathbf{k}}\left(s-s'\right),\\
\chi_{\mathbf{k}}\left(s,s'\right) & =\exp\left(-\tfrac{1}{2\hat{w}}\left(\hat{\tau}_{0}+\mathbf{k}^{2}\right)\left(s-s'\right)\right)\hat{\gamma}_{\mathbf{k}}\left(s+s'\right),
\end{align*}
where $\gamma_{\mathbf{k}}\left(s-s'\right)$ and $\hat{\gamma}_{\mathbf{k}}\left(s+s'\right)$
are arbitrary functions. But in the perturbation expansion there actually
only occur the ``causal'' response functions 
\begin{align}
G_{\mathbf{k}}\left(s_{2},s_{2}';s_{1},s_{1}'\right) & =\theta\left(s_{2}-s_{1}\right)\exp\left(-\frac{1}{w}\left(\tau_{0}+\mathbf{k}^{2}\right)\left(s_{2}-s_{1}\right)\right)\delta\left(\left(s_{2}-s_{1}\right)-\left(s_{2}'-s_{1}'\right)\right),\label{eq:ResponseFunc_ks}\\
\hat{G}_{\mathbf{k}}\left(s_{2},s_{2}';s_{1},s_{1}'\right) & =\theta\left(s_{2}-s_{1}\right)\exp\left(-\frac{1}{\hat{w}}\left(\hat{\tau}_{0}+\mathbf{k}^{2}\right)\left(s_{2}-s_{1}\right)\right)\delta\left(\left(s_{2}-s_{1}\right)+\left(s_{2}'-s_{1}'\right)\right),\nonumber
\end{align}
where $\text{\ensuremath{\left(s_{1},s_{1}'\right)}}$ are the initial
and $\text{\ensuremath{\left(s_{2},s_{2}'\right)}}$ the final length
variables. The delta functions enforce the length variable constraints.
In frequency space the propagators simply are
\begin{align*}
G_{\mathbf{k}}\left(\omega,\omega'\right) & =1/\left(\tau_{0}+\mathbf{k}^{2}-w\left(i\omega+i\omega'\right)\right),\\
\hat{G}_{\mathbf{k}}\left(\omega,\omega'\right) & =1/\left(\hat{\tau}_{0}+\mathbf{k}^{2}-\hat{w}\left(i\omega-i\omega'\right)\right).
\end{align*}
And in fact, the perturbation theory can also be done in the traditional
way by combining segments of trajectories of different types of polymers
and integrating over coordinates, carefully taking account of the
segment length constraints (an example is eq.(\ref{eq:Scaling_function_K_def})
below). The phenomenlogy of the model and the meaning of the relevant
parameters $r_{0}$, $\tau_{0}$ and $\hat{\tau}_{0}$ are examined
after the renormalization group calculation in the next section.

\subsection*{Renormalization group calculation for model A}

\begin{figure}
\centering{}\includegraphics[scale=1.3]{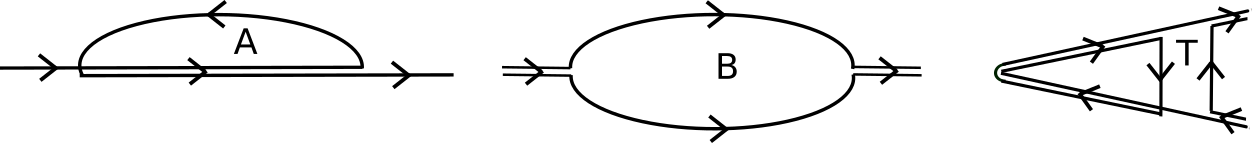}\caption{One loop self energy diagrams, and a two-loop diagram for the composite
operator $\tilde{\psi}\psi$. \label{fig:ModelA1Loop}}
\end{figure}
In this work we restrict ourselves to the case where only single strands
oriented in the same direction form a double strand (model A. Model
B is when only oppositely aligned strands bind, like a sequence ...XYZzyx...).
This means $\hat{g}=0$ and $\chi$ and $\tilde{\chi}$ can be ignored. 

The first step is to examine the scaling behavior of the microscopic
field theory (\ref{eq:FieldTheory}). Because of the symmetries $\varphi\leftrightarrow\tilde{\varphi}$
, $\psi\leftrightarrow\tilde{\psi}$ and $s\leftrightarrow s'$ one
concludes that the model is of order $4$ - there are four different
scaling exponents, say, for $k$, $\omega$, $\varphi$ and $\psi$.
The (naive) scaling exponents are denoted as $\left[k\right]=1$,
$\left[\omega\right]$, $\left[\varphi\right]$ and $\left[\psi\right]$.
The scaling of the wavector (or length) is purely geometric and the
exponent $1$ is exact by definition. Dimensional analysis (the action
must be dimensionless) then requires that a wave vector dimension
also is assigned to the coupling constant $g$. The result of the
linear algebra is $\left[s^{-1}\right]=\left[\omega\right]=2,$ $\left[\varphi\right]=\left[\tilde{\varphi}\right]=3-\epsilon/2$,
$\left[\psi\right]=\left[\tilde{\psi}\right]=4-\epsilon/2$ and $\left[g\right]=\epsilon/2$
as well as $\left[w\right]=0.$ As usual $\epsilon=d-d_{c}$, the
critical dimension $d_{c}$ is $6$. With these scaling exponents
it can already be checked that the excluded volume effect (between
all sorts of polymers) is strongly irrelevant in the renormalization
group sense in the vicinity of $d_{c}=6.$ One final remark is in
order here. $\mathfrak{\mathcal{L}}_{0}$ and $\mathfrak{\mathcal{L}}_{1}$
together provide $5$ linear equations (ignoring the relevant $r_{0}$
and $\tau_{0}$ terms). Four of them already suffice to determine
$\left[\varphi\right]$, $\left[\psi\right],$ $\left[\omega\right]$
and $\left[g\right]$. This is the reason for introducing the ``coupling
constant'' $w$. The physical scaling exponents are written as $\left[f\right]+\eta_{f}$,
for instance $\left[\varphi\right]+\eta_{\varphi}.$ 

The model A perturbation theory generates no one loop contributions
to the $3$-point vertex $g\psi\tilde{\varphi}\tilde{\varphi}$ and
likewise to $g\tilde{\psi}\varphi\varphi$. There only remain the
self energy renormalizations, see fig.(\ref{fig:ModelA1Loop}). The
contributions to the action (\ref{eq:FieldTheory}) are 
\begin{align*}
\left.\Gamma_{\varphi\tilde{\varphi}}^{\left(1\right)}\right|_{k=\mu} & =-g_{0}^{2}\left\{ 2k^{2}I_{1}\left(\epsilon\right)-\left(4w+2\right)i\omega I_{2}\left(\epsilon\right)\right\} ,\\
\left.\Gamma_{\psi\tilde{\psi}}^{\left(1\right)}\right|_{k=\mu} & =-g_{0}^{2}\left\{ 2k^{2}I_{1}\left(\epsilon\right)-2i\left(\omega_{1}+\omega_{2}\right)I_{2}\left(\epsilon\right)\right\} .
\end{align*}
The dimensionless bare coupling constant is $g_{0}=g\mu^{-\epsilon/2},$
where $\mu$ is an arbitrary small wave vector. The integrals $I_{1}$
and $I_{2}$ are defined in appendix A. The mapping to the renormalized
field theory then amounts to the rescalings $\varphi=Z\varphi_{R},$
$\tilde{\varphi}=Z\tilde{\varphi}_{R},$$\psi=Z_{\psi}\psi,$ $\tilde{\psi}=Z_{\psi}\tilde{\psi},$$s=Z_{s}s_{R}$
(only considering the critical point $r_{R}=\tau_{R}=0).$ The formalism
is standard and need not be described in detail here.\cite{Am78}\cite{ZinnJustin}
One result are the flow equations
\begin{align}
\mu\partial_{\mu}g_{R} & =-\frac{\epsilon}{2}g_{R}\left\{ 1-6g_{R}^{2}I_{1}\right\} ,\label{eq:Model_A_Flow}\\
\mu\partial_{\mu}w_{R} & =\frac{\epsilon}{w_{R}}g_{R}^{2}\left(2w_{R}^{2}+w_{R}-1\right)I_{2}\nonumber 
\end{align}
for the renormalized dimensionless coupling constants $g_{R}$ and
$w_{R}.$ For $\mu\rightarrow0$ one finds the infrared stable fixed
point $g_{R}^{2}=\frac{\epsilon}{2}$ and $w_{R}=\frac{1}{2}.$ The
flow equation for $w$ is remarkable, it is nonlinear and has a second
(unphysical) stable fixed point $w_{R}=-1$. The value $\frac{1}{2}$
means that a double strand has a larger size than a single strand
for a given end-to-end length. 

The anomalous contributions to the scaling dimensions at the stable
fixed point to order $O\left(\epsilon\right)$ are 
\begin{equation}
\eta_{\varphi}=\epsilon,\quad\eta_{\psi}=\frac{11}{6}\epsilon,\quad\eta_{\omega}=\frac{5}{3}\epsilon.\label{eq:Model_A_Exponents}
\end{equation}
The scaling exponent for $\omega$ thus is $z=\left[\omega\right]+\eta_{\omega}.$
As a first result one can conclude from the scaling equivalences $\omega\sim k^{z}$
and $s\sim x^{z}$ (where $x$ is a length) that $x\sim s^{1/z}=s^{1/\left(2+5\epsilon/3\right)}.$
Normal diffusive behavior would be $x\sim\sqrt{s}$. A single strand
thus has a much smaller extension. This is plausible - binding to
a double strand as in fig.(\ref{fig:ModelA1Loop}) counteracts diffusion.

\subsection*{Phenomenology of model A}

In this section we attempt to understand the physics at and near the
stable fixed point. The simplest and most important scenario is a
single one-strand polymer. The relevant parameter $r_{0}$ (or its
renormalized counterpart) always is zero. A nonzero value would be
equivalent to a Boltzmann factor for $\varphi$ segments in the partition
sum. However, the polymer is of fixed length and there is no reservoir
of $\varphi$ molecules. This is analogous to first quantized models,
where contributions to $r_{0}$ are generated in perturbation theory,
but the effective $r_{0}$ value is zero.\cite{Cloiz81} 

This is different for the double strand $\psi$. The parameter $\tau_{0}$
(or its renormalized counterpart $\tau$) can be interpreted as energy
per length (a Boltzmann factor in the partition sum) of $\psi$ segments,
and $\varphi$ acts as a reservoir for $\psi$. The parameter $\tau$
in principle can be changed by changing the temperature or the composition
of the solution.

As long as $\tau$ is nonnegative, one would expect that the amount
of double strands $\psi$ is small. In this sense $\psi$ plays the
role of an order parameter. This does not mean that the $\psi$ field
has no effect for $\tau\geq0.$ There could be many short $\psi$
segments with a small statistical weight in the partition sum. This
can be compared with the excluded volume effect, where there also
is no binding energy. The critical exponents depend on the interaction
nevertheless. The region $\tau>0$ in the parameter space of the system
then would be a region of a transition from excluded volume behavior
to model A behavior with exponents (\ref{eq:Model_A_Exponents}).
For $\tau<0$ the binding energy becomes positive and there will be
many $\psi$ segments.

To estimate the amount of double strands with the field theory (\ref{eq:FieldTheory}),
one should introduce an external field $h_{\psi\tilde{\varphi}}\left(\mathbf{x},s\right)$
which couples to $\int ds'\psi\left(\mathbf{x},s,s'\right)\tilde{\varphi}\left(\mathbf{x},s\right)$.
This ``sink'' in principle is a normal $\varphi$ sink, but it describes
the case where a single strand terminates within a double strand,
converting it to a single strand. The one-loop contribution to this
sink looks like graph A in fig.(\ref{fig:ModelA1Loop}) without the
propagator on the r.h.s.

If $u$ denotes the length of the $\psi$ and $v$ the length of the
$\varphi$ propagator, then the total single strand length in in the
loop of graph A of fig.(\ref{fig:ModelA1Loop}) is $s=2u+v$. In $(k,s)$
- space the propagators are (eq.(\ref{eq:ResponseFunc_ks})) $\theta\left(s\right)e^{-k^{2}s}$
and $\theta\left(s\right)e^{-\left(\tau+k^{2}\right)s/w}$ and one
gets
\begin{align}
K\left(\tau,s\right) & =g\int_{0}^{s}dve^{-\tau\frac{s-v}{2w}}\int^{\Lambda}\frac{d^{d}k}{\left(2\pi\right)^{d}}e^{-k^{2}\left(v+\frac{s-v}{2w}\right)}\overset{w=1/2}{=}\frac{g}{\left(2\pi\right)^{\frac{d}{2}}}s^{1-\frac{d}{2}}f\left(\tau s\right),\label{eq:Scaling_function_K_def}\\
f\left(y\right) & =2^{-\frac{d}{2}}\frac{1}{y}\left(1-e^{-y}\right),\nonumber 
\end{align}
where factors $\frac{1}{2}$ originate from the constraint $\delta\left(2u-v-s\right).$
Performing the momentum integral without cutoff $\Lambda$ is allowed
when $\Lambda\sqrt{s}\gg1$. The scaling function $f\left(y\right)$
only depends on $y=\tau s,$ the \emph{negative total binding energy}
of a double polymer of length $s$. 

This leads to a consistent picture. For $y$ large (high temperature,
negative binding energy) $f\left(y\right)\cong2^{-d/2}/y$ and $K\left(\tau,s\right)\sim s^{-d/2}/\tau.$
Therefore the amount of double strands first grows like $1/\tau$
when the temperature is lowered. This growth slows down near $\tau=1/s$
because $f\left(0\right)=const.$ Below the critical temperature (positive
binding energy) $y$ is negative and $f\left(y\right)$ grows exponentially.

The creation of double strands below the critical temperature takes
place in a temperature interval $\tau\sim1/s$ and looks like a first
order phase transition when the chain length $s$ is large. However,
the process is still described by universal critical exponents and
a universal scaling function. Of course, the scaling function and
the critical exponents calculated here only are correct near $d=6$,
but the general picture should be the same in lower dimensions. 

Two somewhat technical remarks are in order. To get a complete response
function one should calculate $\int_{\Lambda^{-2}}^{S}dsK\left(\tau,s\right)$,
the combination of $K\left(\tau,s\right)$ with a $\varphi$ propagator
with length $S-s$ and with $k=0$ (which has value one) on the l.h.s.
of diagram A. This does not change the general picture. The factor
$s^{1-d/2}$ can be understood as the ratio of the volume of the polymer
($\sim s$) to the volume of a box containing the polymer ($\sim\sqrt{s}^{d}$).

\subsection*{Some remarks concerning model B}

\begin{figure}
\centering{}\includegraphics[scale=1.4]{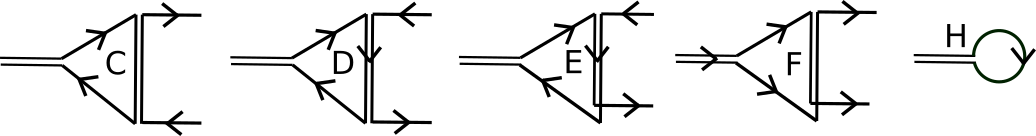}\caption{Some diagrams for model B. Double lines without arrow consist of oppositely
aligned single strands. There also is a hairpin diagram H of order
$g$, which can be inserted into single strand polymers. \label{fig:Model_B_Interaction}}
\end{figure}
In model B oppositely oriented single strands bound to a double strand
do not have an intrinsic direction. In the perturbation theory there
automatically appears a diagram with a $\left\langle \tilde{\chi}\chi\right\rangle $
and a diagram with a $\left\langle \chi\tilde{\chi}\right\rangle $
propagator. Both diagrams contribute the same value.

Model $B$ also allows hairpin configurations. A single polymer strand
can form a loop and then bind to itself to a hairpin configuration,
thus generating a single $\chi$ field. Hairpin configurations are
a well known phenomenon in real DNA physics. 

All this is nothing new in principle, diagram H simply is another
single strand self energy. But the greater number of fields and coupling
constants require more effort. As can be seen from fig.(\ref{fig:Model_B_Interaction})
the perturbation theory now also generates one loop contributions
to the coupling constants.

\subsection*{Conclusions and open problems}

The field theory (\ref{eq:FieldTheory}) defines at least three universality
classes with upper critical dimension $d_{c}=6.$ The high critical
dimension makes it difficult to get accurate values for $d=3.$ But
the qualitative behavior in $d=3$ could be similar nevertheless.
At least this is the case for comparable models with also $d_{c}=6,$
for instance static and dynamic percolation.

To get realistic results for $d=3$ the excluded volume effect also
must be considered. This effect is simple in principle. The Flory
argument gives exact critical exponents in dimensions $1,2$ and $4$,
the critical exponent for $d=3$ is close to the exact value.\cite{ZinnJustin}
A realistic scenario is that the excluded volume interaction only
modifies the critical exponents.

The phenomenology of model A still is not completely clear. The flow
(\ref{eq:Model_A_Flow}) of parameter $w$ to the value $\frac{1}{2}$
in the crossover region is a nonlinear effect, but might still have
a simple explanation. One also should try to calculate more measurable
quantities. For instance, the length of the double strands is determined
by the scaling dimension of the relevant parameter $\tau$ (the temperature).
A two loop diagram contributing to the critical exponent is shown
in fig.(\ref{fig:ModelA1Loop}).

Also open are the model B case and the case of (normal) undirected
polymers, where the internal direction only is a technical device. 

\subsection*{Appendix A: Derivation of the field theory}

The principal idea is to consider directed polymers on the edges of
a finite rectangular lattice. The goal is to count the number of polymer
configurations. To simplify the steps we first consider a simplified
model.

\subsubsection*{A simple model without length variables}

We first assume a Boltzmann factor $p$ for a polymer link. When $k$
enumerates the chain configurations and $l_{k}$ is the number of
links then the partition sum is $Z=\sum_{k}p^{l_{k}}.$

The configurations can be generated with the help of a simple operator
algebra (similar techniques have been used for percolation problems\cite{CS80}).
For each lattice site $i$ there are operators $\bar{a}_{i}$ and
$a_{i}$. The operator $\bar{a}_{i}$ creates the start of a polymer
link at site $i$, the operator $a_{i}$ terminates a link at site
$i.$ All these operators commute and are nilpotent, $a_{i}^{2}=\bar{a}_{i}^{2}=0$.
Nilpotency guarantees that links cannot converge or diverge - there
are no branches and no overlaps. The last property of the operator
algebra is an expectation value $\left\langle \bar{a}_{i}a_{j}\right\rangle _{0}=\delta_{i,j}$.
The expectation value connects a link end to a link start. A formal
expression for the partition sum then is
\[
Z\left(p\right)=\left\langle \prod_{i,j}\left(1+\bar{a}_{i}U_{i,j}a_{j}\right)\right\rangle _{0},
\]
where $U_{i,j}=p$ for next neighbors and $0$ otherwise. The product,
when multiplied out, generates a huge sum. Each combination of next
neighbor links occurs exactly once, but because of the operator algebra
only valid configurations contribute. Pairing $\bar{a}_{i}$ with
$a_{i}$ operators generates chains of $U_{i,j}$ operators, and each
link produces a factor $p$. The partition sum describes the statistical
mechanics of polymer loops. Loops of length two like $U_{1,2}U_{2,1}\left\langle \bar{a}_{1}a_{1}\right\rangle _{0}\left\langle \bar{a}_{2}a_{2}\right\rangle _{0}=p^{2}$
also are allowed. 

For a given $i,j$ pair $1+\bar{a}_{i}U_{i,j}a_{j}=e^{\bar{a}_{i}U_{i,j}a_{j}}$,
and in matrix notation it follows $Z=\left\langle e^{\bar{a}Ua}\right\rangle _{0}.$
A field theory can now be derived with the help of a Hubbard-Stratonovich
transformation
\begin{align*}
Z\left(p\right) & =\left\langle \int\mathscr{\mathcal{D}}\tilde{\varphi}\mathscr{\mathcal{D}}\varphi e^{-\tilde{\varphi}U^{-1}\varphi+\tilde{\varphi}a+\bar{a}\varphi}\right\rangle _{0}=\int\mathscr{\mathcal{D}}\tilde{\varphi}\mathscr{\mathcal{D}}\varphi e^{-\tilde{\varphi}U^{-1}\varphi}\left\langle \prod_{i}\left(1+\tilde{\varphi}_{i}a_{i}\right)\prod_{j}\left(1+\bar{a}_{j}\varphi_{j}\right)\right\rangle _{0}.
\end{align*}
The integrals over $\tilde{\varphi}_{i}$ run along the imaginary
and the integrals over $\varphi_{i}$ along the real axis. The expectation
value can now be evaluated
\begin{equation}
\left\langle ...\right\rangle _{0}=\prod_{i}\left(1+\tilde{\varphi}_{i}\varphi_{i}\right)=e^{\sum_{i}\ln\left(1+\tilde{\varphi}_{i}\varphi_{i}\right)}=e^{\sum_{i}\left\{ \tilde{\varphi}_{i}\varphi_{i}-\frac{1}{2}\left(\tilde{\varphi}_{i}\varphi_{i}\right)^{2}+\frac{1}{3}\left(\tilde{\varphi}_{i}\varphi_{i}\right)^{3}-\frac{1}{4}\left(\tilde{\varphi}_{i}\varphi_{i}\right)^{4}+...\right\} .}\label{eq:ExcludeVolumePhi}
\end{equation}
The first term in the exponent contributes to the propagator, the
second term is the excluded volume interaction. The third and fourth
term are strongly irrelevant in the renormalization group sense but
are kept to render the $\tilde{\varphi}$-integrals finite. To use
this model in a meaningful way one could simply omit all closed loops
in the perturbation theory and add external sources. Perturbation
theory and renormalization group then reproduce the critical exponents
of the $n\rightarrow0$ limit of the $O\left(n\right)$ system.

\subsubsection*{The actual model}

Closed polymer loops disappear automatically when the operators also
have a length index $\mu\in\mathbb{Z}$ which is incremented by $1$
in every link. Furthermore, we anyway need the partition sum of a
polymer of fixed length. This is achieved with the help of commuting
operators $\bar{a}_{i}^{\mu}$ and $a_{i}^{\mu}$ with a site index
$i$ and a length index $\mu$ and the properties $\bar{a}_{i}^{\mu}\bar{a}_{i}^{\nu}=a_{i}^{\mu}a_{i}^{\nu}=0$
and $\left\langle \bar{a}_{i}^{\mu}a_{j}^{\nu}\right\rangle _{0}=\delta_{i,j}\delta_{\mu,\nu}.$
The meaning of the operators is the same except that they now start
or terminate a link with given length index. At a given site two operators
are nilpotent for any length indexes. A formal expression for the
partition sum now is
\[
Z=\left\langle \prod_{i,j,\mu,\nu}\left(1+\bar{a}_{i}^{\mu}U_{i,j}^{\mu,\nu}a_{j}^{\nu}\right)\right\rangle _{0}=\left\langle e^{\bar{a}Ua}\right\rangle _{0}.
\]
The matrix $U_{i,j}^{\mu,\nu}=v_{i,j}\delta_{\mu,\nu+1}$ has value
$1$ for next neighbor sites and incremented length index, otherwise
it has value zero. The algebra is exactly the same except that the
matrix $U$ now increments the length index. Matrix products comprise
$i$ and $\mu$. Since closed loops cannot occur anymore $Z$ actually
is trivial, but this changes when external sources are added. The
transition to a field theory proceeds as above. The harmonic part
of the action becomes $S_{0}^{\left(\varphi\right)}=-\tilde{\varphi}U^{-1}\varphi$,
where the matrix product runs over lattice sites \emph{and} length
indexes. Only the evaluation of the expectation value differs, because
now operators with different length indexes interact,
\begin{align*}
\left\langle e^{\tilde{\varphi}a+\bar{a}\varphi}\right\rangle _{0} & =\left\langle \prod_{i,\mu}\left(1+\tilde{\varphi}_{i}^{\mu}a_{i}^{\mu}\right)\prod_{j,\nu}\left(1+\bar{a}_{j}^{\nu}\varphi_{j}^{\nu}\right)\right\rangle _{0}\\
 & =\left\langle \prod_{i}\left(1+\sum_{\mu}\tilde{\varphi}_{i}^{\mu}a_{i}^{\mu}\right)\left(1+\sum_{\nu}\bar{a}_{i}^{\nu}\varphi_{i}^{\nu}\right)\right\rangle _{0}=\prod_{i}\left(1+\sum_{\mu}\tilde{\varphi}_{i}^{\mu}\varphi_{i}^{\mu}\right).
\end{align*}
The first product over $\mu$ becomes a sum over $\mu$ because of
$a_{i}^{\mu}a_{i}^{\nu}=0$ and likewise the product over $\nu$.
The excluded volume interaction (the last product) now acts between
two polymers strands with arbitrary length index. 

Double polymers are completely identical except that their operators
$\bar{b}_{i}^{\mu,\nu},b_{i}^{\mu,\nu}$ and $\bar{c}_{i}^{\mu,\nu},c_{i}^{\mu,\nu}$
carry two length indexes which are incremented or decremented accordingly
from link to link with matrices
\begin{align}
V_{i,j}^{\mu\nu,\rho\tau} & =v_{i,j}\delta_{\mu,\rho+1}\delta_{\nu,\tau+1},\label{eq:V_W_Matrices}\\
W_{i,j}^{\mu\nu,\rho\tau} & =v_{i,j}\delta_{\mu,\rho+1}\delta_{\nu,\tau-1}.\nonumber 
\end{align}
The expectation value is $\left\langle \bar{b}_{i}^{\mu,\nu}b_{j}^{\rho,\tau}\right\rangle _{0}=\delta_{i,j}\delta_{\mu,\rho}\delta_{\nu,\tau}$
and likewise for $c.$ The matrix $V$ propagates operators $b$ with
single strands aligned parallel, the matrix $W$ propagates operators
$c$ with single strands aligned oppositely. As for the single strand
above the harmonic part of the field theory simply becomes $S_{0}^{\left(\psi,\chi\right)}=-\tilde{\psi}V^{-1}\psi-\tilde{\chi}W^{-1}\chi$,
where the field $\psi$ corresponds to $b$ and the field $\chi$
corresponds to $c$. The matrix products run over the lattice sites
and \emph{two} length indexes.

The model now is complete except for transforming $S_{0}^{\left(\psi,\chi\right)}$
to a continuum (coordinate) representation and except for the interaction
between the different types of polymer strands. The propagator of
the $\psi$ field in $S_{0}^{\left(\psi\right)}$ is the inverse of
the matrix $V$ from eq.(\ref{eq:V_W_Matrices}). The inverse of $V$
can be found by solving the equation $\sum_{j,\rho,\tau}V_{i,j}^{\mu\nu,\rho\tau}f_{j}^{\rho,\tau}=\sum_{j}v_{i,j}f_{j}^{\mu-1,\nu-1}=h_{j}^{\mu,\nu}$
or $\sum_{j}v_{i,j}f_{j}^{\mu,\nu}=h_{j}^{\mu+1,\nu+1}$. The next
neighbor matrix $v_{i,j}$ is diagonal in wave vector space, like
$v\left(k\right)=\left(A+Bk^{2}+...\right)^{-1}.$ The solution of
the equation then is

\[
f^{\mu,\nu}\left(\mathbf{k}\right)=\left(A+Bk^{2}+...\right)\left(1+\partial_{\mu}+\partial_{\nu}+\frac{1}{2}\partial_{\mu}^{2}+...\right)h^{\mu,\nu}\left(\mathbf{k}\right).
\]
This leads to the $\psi$ propagator of the field theory (\ref{eq:FieldTheory})
when the length indexes $\mu,\nu$ are identified with the length
variables $s$ and $s'$. The matrix $W$ for the $\chi$ propagator
has a minus sign for the second length index.

\subsubsection*{Interaction between single and double strands}

The interactions $\mathcal{L}_{1}$ and $\hat{\mathcal{L}}_{1}$ of
eq.(\ref{eq:FieldTheory}) now could be immediately written down,
for completeness the derivation is sketched here. It is assumed that
at any lattice point $i$ only ``diagonal'' operator products $\bar{a}_{i}^{\mu}a_{i}^{\nu}$,
$\bar{b}_{i}^{\mu\nu}b_{i}^{\rho\tau}$ and $\bar{c}_{i}^{\mu\nu}c_{i}^{\rho\tau}$
are different from zero. All other products vanish, $a_{i}^{\mu}a_{i}^{\nu}=a_{i}^{\mu}\bar{c}_{i}^{\nu\rho\tau\gamma}=...=0.$
This naturally leads to the excluded volume interaction (\ref{eq:ExcludeVolumePhi})
between all three types of polymers.

The binding of single strands to double strands is generated by inserting
an additional factor

\begin{align*}
Y & =\prod_{\left(i,j,m\right)\mu\nu}\left(1+g_{1}a_{i}^{\mu}a_{j}^{\nu}\bar{b}_{m}^{\mu\nu}\right)\prod_{\left(i,j,m\right)\mu\nu}\left(1+g_{2}\bar{a}_{i}^{\mu}\bar{a}_{j}^{\nu}b_{m}^{\mu\nu}\right)\\
 & \qquad\prod_{\left(i,j,m\right)\mu\nu}\left(1+g_{3}a_{i}^{\mu}\bar{a}_{j}^{\nu}\bar{c}_{m}^{\mu\nu}\right)\prod_{\left(i,j,m\right)\mu\nu}\left(1+g_{4}\bar{a}_{i}^{\mu}a_{j}^{\nu}c_{m}^{\mu\nu}\right)
\end{align*}
into the expectation value of the partition sum. The constants $g_{1},g_{2}$
are weights for transitions between two single strands $\varphi$
and a double strand $\psi$, the constants $g_{3},g_{4}$ analogously
for a double strand $\chi$. The products run over all length indexes
$\mu,\nu$ and all triangles of next neighbor lattice points $\left(i,j,m\right)$,
with $m$ not in the middle. All terms in $Y$ are quasi local. The
length indexes are not arbitrary. In the $g_{3}$ term the index $\mu$
of an \emph{incoming} single strand becomes the first index of a double
strand, in the $g_{4}$ term the first index $\mu$ of a double strand
becomes the index of an \emph{outgoing} single strand, in accord with
the fact that the matrix $W$ (the $\chi$ propagator) increments
the first length index. When the Hubbard transformations have been
performed there remains the expectation value 
\[
\left\langle Y\prod\left(1+\tilde{\varphi}a\right)\prod\left(1+\bar{a}\varphi\right)\prod\left(1+\tilde{\psi}b\right)\prod\left(1+\bar{b}\psi\right)\prod\left(1+\tilde{\chi}c\right)\prod\left(1+\bar{c}\chi\right)\right\rangle _{0}.
\]
Contraction of the operators of $Y$ among themselfes only produces
uninteresting constants. Mixed contractions produce local terms which
are irrelevant or relevant and present in the model anyway. Contraction
of $Y$ only with the other factors leads to the interactions $\mathcal{L}_{1}$
and $\hat{\mathcal{L}}_{1}$.

\subsection*{Appendix B: One loop integrals}

Required are the $k^{2}$, $i\omega$ and $i\omega_{1}+i\omega_{2}$
parts of the integrals 
\begin{align*}
J_{1}= & \frac{1}{K_{d}\left(2\pi\right)^{d}}\int\frac{d^{d}p}{\left(\tau_{0}+\left(p+k\right)^{2}-2wi\omega\right)\left(r_{0}+p^{2}-i\omega\right)},\\
J_{2}= & \frac{1}{K_{d}\left(2\pi\right)^{d}}\int\frac{d^{d}p}{\left(r_{0}+\left(p+k\right)^{2}-i\omega_{1}\right)\left(r_{0}+p^{2}-i\omega_{2}\right)}
\end{align*}
for $d=6-\epsilon$ and $\left|k\right|=\mu$, calculated with dimensional
regularization. The constant $K_{d}$ is $K_{d}=2^{-d+1}\pi^{-d/2}/\Gamma\left(d/2\right).$
The $i\omega$ contributions come from
\[
I_{2}=\frac{1}{K_{d}\left(2\pi\right)^{d}}\int\frac{d^{d}p}{\left(p+k\right)^{2}\left(p^{2}\right)^{2}}=\frac{\mu^{-\epsilon}}{\epsilon}+...
\]
The $k^{2}$ contributions come from
\[
I_{1}=-\frac{\partial^{2}}{\partial k^{2}}\frac{1}{K_{d}\left(2\pi\right)^{d}}\int\frac{d^{d}p}{\left(p+k\right)^{2}p^{2}}=\frac{\mu^{-\epsilon}}{3\epsilon}\left(1-\frac{\epsilon}{4}+...\right).
\]

\bigskip{}
\bigskip{}
\bigskip{}

\bigskip{}


\begin{thebibliography}{1}

\bibitem{DeGen72}P.G. de Gennes. Exponents for the excluded volume problem as derived by the Wilson method. \textit{Phys. Lett. A}, 38:339-340, 1972.
\bibitem{Edwards1965}S. F. Edwards. The statistical mechanics of polymers with excluded volume. \textit{Proc. Phys. Soc.}, 85:613-624, 1965.
\bibitem{Cloiz81}J. des Cloizeaux. Polymers in solutions: principles and applications of a direct renormalization method. \textit{J. Physique}, 42:635, 1981.
\bibitem{ZinnJustin}J. Zinn-Justin. \textit{Quantum Field Theory and Critical Phenomena.} Clarendon Press, 1996.
\bibitem{KMP00}Y. Kafri, D. Mukamel, L. Peliti. Why is the DNA Denaturation Transition First Order? \textit{Phys. Rev. Lett.}, 85, No 23:4988-4991, 2000.
\bibitem{VK2017}A. Vologodskii, MD Frank-Kamenetskii. DNA melting and energetics of the double helix. \textit{Phys Life Rev}, 1-43, 2017.
\bibitem{Am78}D.J. Amit. \textit{Field Theory, the Renormalization Group and Critical Phenomena.} McGRAW-HILL, 1978.
\bibitem{CS80}J.L. Cardy and R.L. Sugar. Directed percolation and Reggeon field theory. \textit{J. Phys.}, A13:L423-L427, 1980.

\end{thebibliography}
\end{document}